 \documentclass[final,5p,times,twocolumn]{elsarticle}

\usepackage{amssymb}
\usepackage{multirow}
\journal{Chemical Physics Letters}

\begin{document}

\begin{frontmatter}

%% Title, authors and addresses

%% use the tnoteref command within \title for footnotes;
%% use the tnotetext command for the associated footnote;
%% use the fnref command within \author or \address for footnotes;
%% use the fntext command for the associated footnote;
%% use the corref command within \author for corresponding author footnotes;
%% use the cortext command for the associated footnote;
%% use the ead command for the email address,
%% and the form \ead[url] for the home page:
%%
%% \title{Title\tnoteref{label1}}
%% \tnotetext[label1]{}
%% \author{Name\corref{cor1}\fnref{label2}}
%% \ead{email address}
%% \ead[url]{home page}
%% \fntext[label2]{}
%% \cortext[cor1]{}
%% \address{Address\fnref{label3}}
%% \fntext[label3]{}

\title{Anion-radical oxygen centers in small (AgO)$\mathrm{_n}$ clusters: density functional theory predictions}

%% use optional labels to link authors explicitly to addresses:
%% \author[label1,label2]{<author name>}
%% \address[label1]{<address>}
%% \address[label2]{<address>}

\author[label1,label2]{Egor V. Trushin}

\author[label1,label3]{Igor L. Zilberberg\corref{cor1}}

\cortext[cor1]{Corresponding author, E-mail: I.L.Zilberberg@catalysis.ru}

\address[label1]{Boreskov Institute of Catalysis, Siberian Branch of the Russian Academy of Sciences, pr. Lavrentieva 5, 630090, Novosibirsk, Russia}

\address[label2]{Friedrich-Alexander-Universit\"{a}t Erlangen-N\"{u}rnberg, Lehrstuhl f\"{u}r Theoretische Chemie, Egerlandstra\ss e 3,
91058, Erlangen, Germany}

\address[label3]{Novosibirsk State University, ul. Pirogova 2, 630090, Novosibirsk, Russia}

\begin{abstract}
Anion-radical form of the oxygen centers O$^{-}$ is predicted at the DFT level for small silver oxide particles having the AgO stoichiometry. Model clusters (AgO)$_\mathrm{n}$ appear to be ferromagnetic with appreciable spin density at the oxygen centers. In contrast to these clusters, the Ag$_2$O model cluster have no unpaired electrons in the ground state. The increased O/Ag ratio in the oxide particles is proved to be responsible for the spin density at oxygen centers.
\end{abstract}


\end{frontmatter}

\section{Introduction}
\label{}

In the last decades silver oxide has been actively studied  in numerous works due to its importance in various applications. In particular, silver oxide is considered to be used in batteries \cite{Pan2007}, molecular sensors \cite{Tominaga2003} and bactericidal materials \cite{Dellasega2008, Wang2010}. The interaction of oxygen with metal silver surface leads to the formation of silver oxide overlayers, which play a crucial role in the silver-catalyzed ethylene epoxidation and the partial oxidation of methanol to formaldehyde \cite{Li2003, Schnadt2009}.

Photoelectron spectrum for in-situ prepared AgO bulk samples contains a double-peak O1s signal with a remarkable chemical shift of 2.9 eV between the two components \cite{Bielmann2002}. The authors assigned the splitting of O1s peak via the co-existence of  two nonequivalent oxygen in -1 and -2 oxidation states in silver oxide structure. Later, it was also remarked that silver oxide thin films with high oxygen content contain the same oxygen forms \cite{Kaspar2010}. This is in agreement with earlier DFT calculations for the AgO oxide in which allowed the authors to suggest appearance of "holes" in electronic structure of O$^{2-}$ oxygen \cite{Park1994}. 

The silver-catalyzed ethylene epoxidation reaction is commonly considered to be determined by the competition between two oxygen forms with different O1s electron binding energies called electrophilic and nucleophilic oxygen species \cite{Grant1985}. The electrophilic oxygen is suggested to be responsible for the epoxidation route of ethylene oxidation, while the nucleophilic oxygen governs the combustion \cite{Santen1987}. For a long time the nature of electrophilic oxygen has been debated in terms of atomic \cite{Bukhtiyarov2001} and molecular \cite{Boronin1999} forms. Although, in a number of works the nucleophilic and electrophilic oxygen species are proposed to be two atomically adsorbed forms with different charge states, both forms were usually considered as the O$^{2-}$ species with closed shells. The first model of radical oxygen, to the best of our knowledge, belongs to Carter and Goddard \cite{Carter1988, Carter1989}, who introduced so-called surface atomic oxyradical oxygen (SAO) O$^{-}$ in addition to the standard oxide oxygen O$^{2-}$ on base of a cluster Ag$_3$O.  Later, the increase of electrophilic atomic oxygen concentration was experimentally shown to accelerate the ethylene epoxidation reaction, while increasing the concentration of nucleophilic oxygen slows it down \cite{Bukhtiyarov2006}. 

The origin of atomic electrophilic oxygen on silver catalyst surface and its electronic state are still not completely understood. The electrophilic oxygen is known to be formed on the silver surface at saturation with oxygen accompanied by the formation of silver oxide overlayers \cite{Li2003} and has O1s binding energy about 2 eV higher than that for the nucleophilic oxygen. This oxygen also is characterized by a lower effective charge - about 1 a.u. \cite{Kaichev2003}. 

In our previous paper \cite{Ruzankin2004} closed and open shell states of oxygen adsorbed on silver were simulated based on the Ag$_2$O molecule to model nucleophilic and electrophilic species, respectively. Unlike Carter-Goddard SAO in Ag$_3$O model having single unpaired electron at the oxygen center, the open-shell Ag$_2$O was predicted to be of the singlet biradical type having two electrons with antiparallel spins separated between oxygen and silver. The increase of the O1s energy and disappearence of the pre-edge feature of O$_\mathrm{K}$-edge XAS spectrum for the electrophilic oxygen were in a perfect agreement with the experimental observations. Since the open-shell oxygen form emerged in the excited state of the system, the question arises for which oxo silver systems the atomic electrophilic oxygen appears in the ground state.

It is well-known that nanoscale transition metal oxides possesses unusual structural, electronic, magnetic and catalytic properties, which are very different from the same properties of bulk samples \cite{Cox2010}. Nanostructured silver oxide is not an exception revealing extreme photoactivity \cite{Peyser2001}. However, reliable experimental data on the atomic structure of silver oxide clusters are currently lacking. As a rule, quantum chemical calculations up to now have been focused on the study of silver oxide clusters with regular stoichiometry Ag$_2$O \cite{Burgel2006} or oxygen adsorption on silver clusters  \cite{BK1999, Schmidt2003, Klacar2010}.Silver oxide clusters with the stoichiometry AgO has not yet been studied theoretically to the best of our knowledge.

\begin{figure}[h]
\begin{center}
\includegraphics[width=0.8\linewidth]{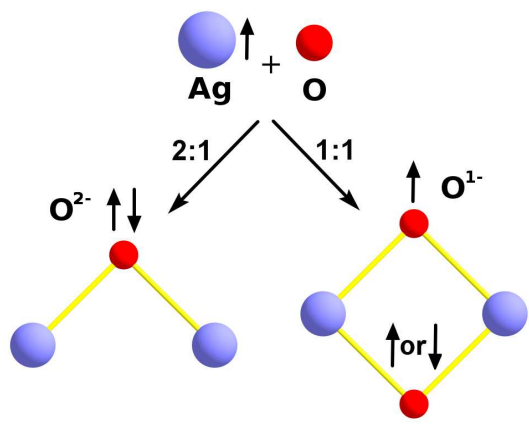}
\caption{The scheme, which illustrates the appearance of oxygen species O$^{2-}$ and O$^{-}$ due to the interaction between silver and oxygen with stoichiometry Ag$_2$O and AgO, respectively.}
\label{ris:Sheme}
\end{center}
\end{figure}

Nanostructured silver oxide AgO can have magnetic properties due to the appearence of anion-radical oxygen in the structure which is an obvious result of a deficiency of the 1s(Ag) electrons to form the O$^{2-}$ centers. In contrast to the interaction between silver and oxygen in stoichiometry Ag$_2$O, in AgO each oxygen atom is able to withdraw only one silver 5s electron giving rise to the oxygen state O$^{-}$. Formation of the O$^{-}$ state in AgO is schematically illustrated in figure \ref{ris:Sheme}. Although mixed oxidation state of silver atoms might be also realized, it is not the case for small AgO clusters as will be shown below.

In case of silver oxide AgO unpaired electrons are to be localized on oxygen atoms, contrary to the majority of the magnetic transition metal oxide systems. Although magnetism of small AgO clusters has not been experimentally proven yet, isoelectronic copper oxide nanoparticles are known to exhibit ferromagnetic or paramagnetic properties which is explained by the increased O to Cu stoichiometric ratio \cite{Yermakov2007, Yang2010}.

In the present work, the existence of anion-radical form of oxygen centers in the ground state is demonstrated for the model silver oxide clusters (AgO)$\mathrm{_n}$ (n = 1-4, 6). The properties of oxygen in the clusters (AgO)$\mathrm{_n}$ are compared with the properties of oxygen in simple silver oxide clusters having various stoichiometries to show how an increase in the relative content of oxygen leads to an increase in spin density at oxo center. The electrophilic oxygen centers on silver responsible for the epoxidation is suggested to have anion-radical structure found for the small (AgO)$_\mathrm{n}$ clusters.

\section{Methods}
All quantum chemical calculations in the present work were carried out within the density functional theory using the Amsterdam Density Functional package \cite{Velde2001, FonsecaGuerra1998}. The combination of the exchange functional Perdew and Wang 1986 (PW86x) \cite{Perdew1986} and Perdew and Wang 1991 correlation functional (PW91c) \cite{Perdew1992} has been applied using polarized triple-zeta basis set of Slater-type orbitals (TZP). To take into account relativistic effects ZORA \cite{Lenthe1999} was additionally used for silver atoms.

Geometry optimizations of all clusters were performed, starting from structures having reasonable Ag-O bond lengths. Optimizations were carried out without restriction of symmetry. The optimized geometries appear to have positive vibrational frequencies which means that these structures correspond to local minimums on the potential energy surface. For all structures the states with different spin projections S$_z$ were checked in spin-polarized calculations. 

\section{Results and discussions}

\begin{figure*}[t]
\includegraphics[width=1\linewidth]{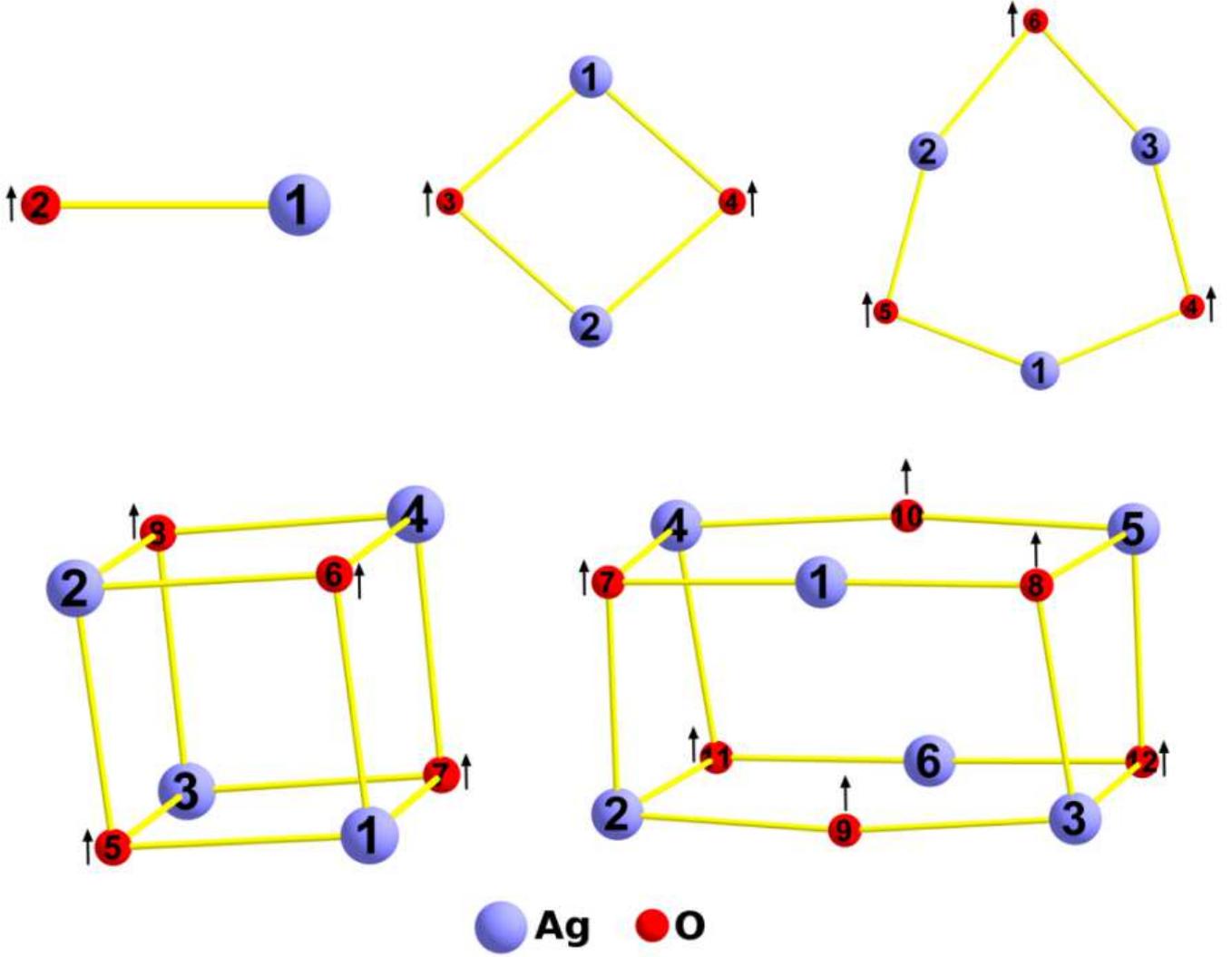}
\caption{The structure of the clusters (AgO)$\mathrm{_n}$ (n = 1-4, 6) as revealed at the PW86-PW91/ZORA-TZP level. Arrows point out the localization of unpaired electrons in the ground state (ferromagntic) of clusters.}
\label{ris:AgOgeoms}
\end{figure*}

The smallest silver oxide cluster AgO has calculated bond length of 2.06 \AA, which is in a good agreement with the experimental value of 2 \AA ~\cite{Huber1979}. The structure of (AgO)$_2$ is a rhombus with the Ag-Ag distance of 2.91 \AA, the O-O distance of 3.27 \AA, and the Ag-O distance 2.19 \AA. Experimental bond lengths for Ag$_2$ and O$_2$  molecules are 2.69 \AA ~ and 1.21 \AA, respectively. Thus, both the Ag-Ag and O-O bonds in (AgO)$_2$ are weaker than their homonuclear counterparts. The (AgO)$_3$ cluster has planar ring-like structure. In thus cluster the Ag-Ag bond is significantly smaller than the O-O bond, and, consequently, the structure of this cluster is not a perfect hexagon. The average Ag-O bond length in (AgO)$_3$ is 2.14 \AA, being close enough to that value for (AgO)$_2$. The distorted cubic structure is realized for the cluster (AgO)$_{4}$ containing the cluster (AgO)$_{2}$ as a building block. The average Ag-O bond length is 2.31 \AA, being markedly larger than that in smaller clusters. The Ag-Ag and O-O distances of 3.31 \AA ~ and 3.22 \AA, respectively, become close to each other. The structure of (AgO)$_{6}$ is simply an extension of the (AgO)$_{4}$ geometry with the addition of a (AgO)$_2$ unit with the formation of distorted prism. Although the Ag-O bond lengths vary between 2.16 and 2.34 \AA, the average Ag-Ag and O-O distances remain close to the corresponding ones in  (AgO)$_{4}$. Optimized geometries of the clusters (AgO)$\mathrm{_n}$ (n = 1-4, 6) in the ground states obtained in the calculations are shown in figure \ref{ris:AgOgeoms}.

\begin{figure*}[t]
\includegraphics[width=1\linewidth]{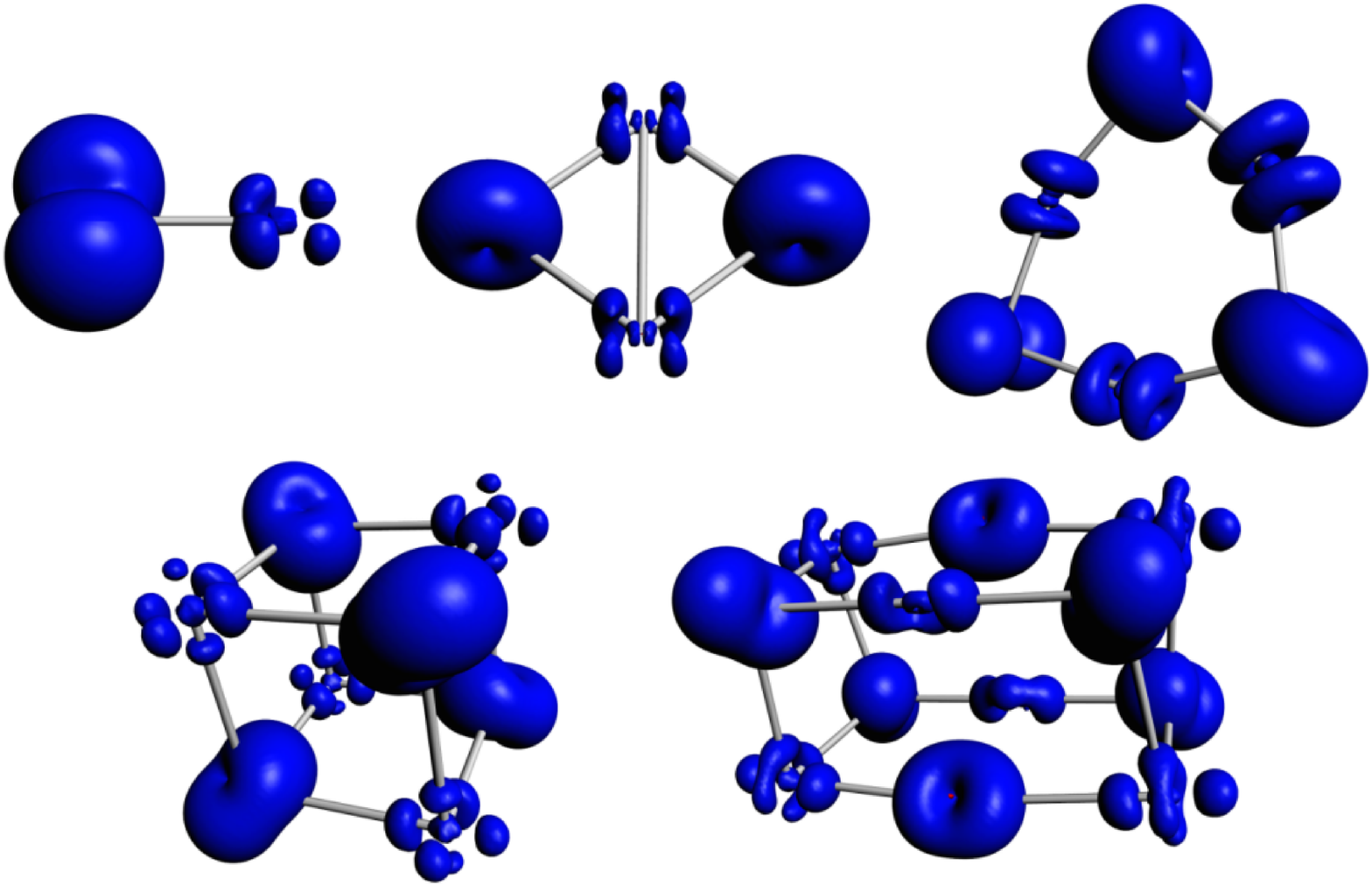}
\caption{Spin density distribution for the clusters (AgO)$\mathrm{_n}$ (n = 1 - 4, 6) in ground states as revealed at the PW86-PW91/ZORA-TZP level}
\label{ris:AgOspindens}
\end{figure*}

Most metal oxides are predominantly ionic in its structural features \cite{Cox2010}. So, calculated small silver oxide clusters (AgO)$_{\mathrm{n}}$ have the same structural motifs as some other well-known ionic clusters, for instance, NaCl, MgO, CsF, etc \cite{Johnston2002, Nayak1998}. Ag-O bond lengths in (AgO)$\mathrm{_n}$ with n=1-3 are in magnitude the same as in different bulk silver oxides \cite{Allen2011}. Ag-O distances in (AgO)$_{4}$ and (AgO)$_{6}$ are similar to the Ag-O bond lengths in silver oxide thin films or oxygen adsorbed on silver surface \cite{Wang2002}.

The stability of clusters (AgO)$\mathrm{_n}$ with n $>$ 6 was estimated by the value of the atomization energy (E$\mathrm{_a}$) defined as

$$ \mathrm{E_a (n) = E(Ag) + E(O) - \frac{1}{n}E_n} $$

where E(Ag), E(O) are the total energies of Ag and O atoms, and E$_\mathrm{n}$ is the total energy of the cluster (AgO)$_\mathrm{n}$.

In this definition, E$\mathrm{_a}$ is positive for clusters that are stable with respect to the complete atomization. The value of E$\mathrm{_a}$ increases monotonically with increasing size n, implying the possibility of larger clusters formation (table \ref{tabular:AgO}). For instance, in our work \cite{Trushin2012} dependence of atomization energy of (ZnO)$\mathrm{_n}$ (n=2-9) of the size also increased monotonically and larger clusters have been observed experimentally \cite{Bulgakov}.

\renewcommand{\arraystretch}{1}
\begin{table*}[t]
\caption{The size of cluster n, the atomization energies E$\mathrm{_a}$, spin-projection S$_z$, mean value for the spin-squared operator $\langle$S$^2 \rangle$, Mulliken charges $\rho$ and spin densities $\mathrm{\rho_{s}}$ for the ground (high-spin) states of clusters, as well as those for excited (low-spin) clusters.}
\label{tabular:AgO}
\begin{center}
\begin{small}
\begin{tabular}{l@{\hspace{0.03cm}}|c|*{11}{@{\hspace{0.03cm}}c@{\hspace{0.03cm}}|}@{\hspace{0.03cm}}c}
\hline \hline
 \multicolumn{3}{c|}{n} & 1 & 2 & 3 & 3 & 4 & 4 & 4 & 6 & 6 & 6 & 6 \\
\hline
 \multicolumn{3}{c|}{ S$_z$} &0.50 & 1 & 1.50 & 0.5 & 2 & 1 & 0 & 3 & 2 & 1 & 0 \\
\hline
 \multicolumn{3}{c|}{$\langle$S$^2 \rangle$} &0.76 & 2.02 & 3.77 & 1.77 & 6.03 & 3.01 & 1.99 & 12.03 & 6.13 & 2.76 & 1.82 \\
\hline
 \multicolumn{3}{c|}{ E$\mathrm{_a}$, eV} &4.227 & 5.317 & 5.814 & 5.780 & 5.855 & 5.771 & 5.802 & 6.158 & 6.146 & 6.145 & 6.144 \\
\hline 
 \multicolumn{13}{c}{}\\[-9.5pt]
 \hline
\multicolumn{3}{c|}{i*} & \multicolumn{10}{c}{Mulliken charges $\rho$ on atoms}\\
\hline
\multicolumn{3}{c|}{1} & 0.547 & 0.629 & 0.689 & 0.690 & 0.656 & 0.662 & 0.653 & 0.775 & 0.869 & 0.923 & 0.888 \\
\multicolumn{3}{c|}{2} & -0.547 & 0.629 & 0.625 & 0.618 & 0.656 & 0.691 & 0.653 & 0.648 & 0.646 & 0.633 & 0.627 \\
\multicolumn{3}{c|}{3} & & -0.629 & 0.625 & 0.624 & 0.656 & 0.663 & 0.669 & 0.649 & 0.649 & 0.636 & 0.629 \\
\multicolumn{3}{c|}{4} & & -0.629 & -0.662  &  -0.680 & 0.656 & 0.668 & 0.670 & 0.648 & 0.648 & 0.609 & 0.619 \\
\multicolumn{3}{c|}{5} & & & -0.661 &  -0.671 & -0.656 & -0.683 & -0.645 & 0.648 & 0.648 & 0.607 & 0.620 \\
\multicolumn{3}{c|}{6} & & & -0.615  & -0.582 &  -0.656 & -0.669 &  -0.645 & 0.786 & 0.888 & 0.928 & 0.942 \\
\multicolumn{3}{c|}{7} & & & & & -0.656 &  -0.662 & -0.678  & -0.700 & -0.731 & -0.743 & -0.746 \\
\multicolumn{3}{c|}{8} & & &  & & -0.656 & -0.669 & -0.677 & -0.699 & -0.732 & -0.742 & -0.741 \\
\multicolumn{3}{c|}{9} & & & & & & &  & -0.679 & -0.714 & -0.680 & -0.674 \\
\multicolumn{3}{c|}{10} & & & & & & &  & -0.679 & -0.716 & -0.686 & -0.689 \\
\multicolumn{3}{c|}{11} & & & & & & & & -0.697 & -0.727 & -0.743 & -0.739 \\
\multicolumn{3}{c|}{12} & & & & & & & & -0.698 & -0.729 & -0.742 & -0.736 \\
\hline
 \multicolumn{13}{c}{}\\[-9.5pt]
 \hline
\multicolumn{3}{c|}{i*} & \multicolumn{10}{c}{Mulliken spin densities $\rho_\mathrm{s}$ on atoms}\\
\hline
\multicolumn{3}{c|}{1} &-0.030 & 0.065 & 0.149 & 0.151 & 0.109 & -0.032 & -0.001 & 0.299 & 0.145 & 0.005 & 0.024 \\
\multicolumn{3}{c|}{2} & 1.030 & 0.065 & 0.131 & 0.057 & 0.109 & 0.117 & -0.002 & 0.108 & 0.117 & 0.094 & 0.041 \\
\multicolumn{3}{c|}{3} & & 0.934 & 0.124 & 0.051 & 0.109 & 0.018 & 0.121 & 0.109 & 0.118 & 0.097 & 0.014 \\
\multicolumn{3}{c|}{4} & & 0.934 & 0.878 & 0.827 & 0.109 & 0.112 & -0.122 & 0.108 & 0.118 & 0.063 & 0.001 \\
\multicolumn{3}{c|}{5} & & & 0.870 & 0.830 & 0.890 & 0.239 & 0.787 & 0.108 & 0.117 & 0.063 & -0.014 \\
\multicolumn{3}{c|}{6} & & & 0.848 & -0.917 & 0.890 & 0.865 & 0.791 & 0.311 & 0.121 & 0.006 & 0.013 \\
\multicolumn{3}{c|}{7} & &  &  &  & 0.890 & -0.181 & 0.870 & 0.010 & 0.582 & 0.354 & 0.066 \\
\multicolumn{3}{c|}{8} & & & &  & 0.890 & 0.862 & -0.871 & -0.005 & 0.585 & 0.376 & -0.216 \\
\multicolumn{3}{c|}{9} & & & &  &  & &  & 0.893 & 0.486 & -0.452 & 0.779 \\
\multicolumn{3}{c|}{10} & & & & &  & & & 0.893 & 0.477 & 0.653 & -0.478 \\
\multicolumn{3}{c|}{11} & & & & & & & & 0.797 & 0.569 & 0.361 & -0.072 \\
\multicolumn{3}{c|}{12} & & & & & & & & 0.795 & 0.563 & 0.378 & -0.158 \\
\hline \hline
\multicolumn{14}{l}{\scriptsize *number of atom in cluster (see fig. \ref{ris:AgOgeoms})}
\end{tabular}
\end{small}
\end{center}
\end{table*}

Considered (AgO)$\mathrm{_n}$ clusters appear to have an electron structure in which unpaired electrons are localized almost exclusively on oxygen centers. Spins on oxygen are coupled through non-magnetic silver centers to form a ferromagnetic structure. Such spin ordering differs from that for magnetic oxides in which spins are localized on metal centers while the oxygen centers serve as non-magnetic bridges as shown in particular in previous computational studies on the magnetic properties of transition metal oxides clusters \cite{Nayak1998, Reddy1999, Wang2010a}. Figure \ref{ris:AgOspindens} shows spin density distribution for the (AgO)$\mathrm{_n}$ clusters in ground states.

The low-spin states of the clusters (AgO)$\mathrm{_n}$ under consideration are less stable than the high spin states. These states might be assigned to antiferromagnetic ordering of spins on oxygen centers. Such magnetic structures are given by broken-symmetry solutions revealed by the distinct large value of the spin contamination, i.e. the excess of the mean value for the spin-squared operator over that for the given value of the spin projection S$_z$(S$_z$+1) \cite{Zilberberg2004, Zilberberg2004a}. The spin contamination in DFT corresponds to the negative spin density or the spin polarization as was shown by Wang, Becke and Smyth \cite{Wang1995}. In DFT, the status of spin contamination is quite different from that of the Hartree-Fock theory (where it is an indicator of the spin-symmetry purity) since the Kohn-Sham (KS) determinant is not a correct wave function and so could not be contaminated in the sense of Hartree-Fock theory. The spin contamination calculated for the KS determinant indicates in what extent the KS orbitals are polarized. It is worthwhile noting that KS orbitals are always polarized (unrestricted) for open-shell molecular systems as the effective potentials for the one-electron equations differ between $\alpha$ and $\beta$ electrons as was pointed out by Pople and co-authors \cite{Pople1995}. The spin contamination in DFT might be explored as the effective number of spatially separated electrons with antiparallel spins: if the spin contamination is about 1.0 then there is one such pair, 2.0 - two pairs, etc \cite{Zilberberg2004}. The exception is the Ag$_6$O$_6$ cluster in the quintet state, in which  there are four unpaired electrons having parallel spins almost equally delocalized over six oxygen atoms. (AgO)$_{2}$ cluster has not a singlet state with a rhombic atomic geometry.

In the case of the cluster (AgO)$_6$ ferromagnetic and antiferromagnetic states are almost degenerate. In the different spin states system has virtually the same geometry with a slightly different bond lengths. This indicates that the unpaired electrons in this system interact weakly with each other. One may assume that with an increase in cluster size, the low-spin antiferromagnetic state is to become energetically more favourable than high-spin ferromagnetic state.

\begin{table*}[t]
\caption{Mulliken charge $\rho$ and spin density $\rho_{\mathrm{s}}$ on atoms and energies of O$\mathrm{_{1s}}$ Kohn-Sham orbitals (for $\alpha$ and $\beta$ sets - in spin-polarized cases) of the clusters Ag$_2$O, Ag$_3$O$_2$, AgO and AgO$_2$}
\label{tabular:2}
\begin{center}
\begin{tabular}{c|c|c|c|c|c|c|c}
\hline
 & \multirow{2}*{S$^2$} & \multirow{2}*{S$_z$} & \multicolumn{2}{c|}{$\mathrm{\rho}$} &  \multicolumn{2}{c|}{$\mathrm{\rho_{s}}$} & \multirow{2}*{$\mathrm{E_{KS}(O_{1s})}$ ($\alpha$;$\beta$), eV} \\
\cline{4-7} 
 & & & Ag & O & Ag & O & \\
 \hline
Ag$_2$O & 0 & 0 &  0.41 &  -0.82 & 0 & 0 & -509.54\\
Ag$_3$O$_2$ & 0.79 & 0.5 & 0.49 & -0.72 & -0.17 & 0.52 & -509.85; -509.50 \\
AgO & 0.76 & 1 & 0.54  & -0.54 & -0.03  & 1.03 & -511.42; -510.89 \\
AgO$_2$ & 3.76 & 1.5 & 0.77  & -0.38 & 0.23  & 1.38 & -513.52; -512.75 \\
\hline
\end{tabular}
\end{center}
\end{table*}

Thus, in the clusters with stoichiometry AgO in both the high-spin and low-spin states the oxygen centers appear to be of the anion-radical type O$^-$. With the increase in the cluster size the charge and spin density on oxygen is reduced reflecting gradual shift of the electron density to silver.

One can expect that bonding mechanism in the (AgO)$\mathrm{_n}$ clusters is predominantly ionic and is determined by electron-pair formed between 5s electron of Ag and 2p electron of O. However, the anion-radical oxygen has unpaired electron on itself implying that 3-electron bonding can be realized. In order to better understand bonding mechanism in the (AgO)$\mathrm{_n}$ model systems bond energy analysis available in the ADF program \cite{Bickelhaupt1998, Bickelhaupt2007} is applied. The interaction energy $\Delta$E$\mathrm{_{int}}$ are decomposed into three physically meaningful terms: 
$$\Delta E \mathrm{_{int}} = \Delta V \mathrm{_{elst}} + \Delta E \mathrm{_{Pauli}} + \Delta E \mathrm{_{oi}}$$
which correspond to the classical electrostatic interaction energy, the Pauli repulsion energy and orbital interaction energy, respectively. The 3-electron bonding energy, which can be determined as $\Delta$E$\mathrm{_{Pauli}}$ + $\Delta$E$\mathrm{_{oi}}$, contributes about 20-30\% to the total bonding energy of (AgO)$\mathrm{_n}$ systems, while the classical electrostatic interaction energy makes up 70-80\% of the total bonding energy. It is the 3-electron bonding that is responsible for the stabilization of the anion-radical oxygen in (AgO)$\mathrm{_n}$ clusters.

\begin{figure}[h]
\includegraphics[width=1\linewidth]{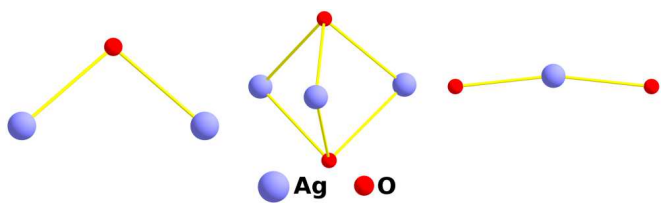}
\caption{The structure of the clusters Ag$_2$O, Ag$_3$O$_2$ и AgO$_2$ as revealed at the PW86-PW91/ZORA-TZP level}
\label{ris:AgnOmgeoms}
\end{figure}

To compare the properties of oxygen in (AgO)$\mathrm{_n}$ clusters with those of the clusters with other stoichiometries, the Ag$_2$O, Ag$_3$O$_2$ and AgO$_2$ systems were calculated. Their structure is shown in figure \ref{ris:AgnOmgeoms}. Table \ref{tabular:2} shows the values of the spin projection S$_z$, mean value of spin-squared operator $\langle$S$^2 \rangle$, Mulliken charges $\mathrm{\rho}$ and spin densities $\mathrm{\rho_{s}}$ on atoms and energies of O1s Kohn-Sham orbitals for $\alpha$ and $\beta$ sets.

With the oxygen stoiciometry increasing in the sequence among Ag$_2$O, Ag$_3$O$_2$, AgO and AgO$_2$ clusters the negative charge on oxygen decreases causing the drop of the O1s Kohn-Sham levels. This finding agrees well with the experimental data by Kaichev with co-authors who suggested that the charge on oxygen is responsible for the relative value of the binding energy of oxygen O1s electrons on the silver surface \cite{Kaichev2003}: the increase in the charge on oxygen decreases the binding energy of its 1s electron.

Oxygen in Ag$_2$O cluster having zero spin density and a significant negative charge is similar in its properties to the oxygen in bulk Ag$_2$O silver oxide with formal oxidation state of -2. The same properties are inherent to atomic nucleophilic oxygen species which are routinely registered during adsorption of oxygen on regular silver surface.

In turn, anion-radical oxygen appearing in small (AgO)$\mathrm{_n}$ clusters has significant spin density and smaller magnitude of negative charge. The binding energy of the O1s level of the anion-radical oxygen is significantly stabilized as compared to that of the oxygen in Ag$_2$O cluster. All these oxygen parameters seem to correspond to the oxidation state of -1.

It is O$^{-}$ oxygen that was suggested to reveal itself in the photoelectron spectroscopy of bulk silver oxide AgO. At a high coverage oxygen may intiate a reconstruction of silver surface into AgO-like thin oxide film. It may cause the formation of anion-radical oxygen center on surface which may govern silver-catalyzed ethylene epoxidation. The latter suggestion is supported by the fact that the electron acceptors such as chlorine facilitate the ehylene epoxidation reaction (allowing the formation of electrophilic oxygen at low oxygen concentrations), while the electron donors such as cesium slow down the reaction \cite{Jankowiak2005}.

\section{Conclusions}

In this Letter, we performed density functional study on small silver oxide clusters (AgO)$\mathrm{_n}$ (n=1-4, 6) to demonstrate the possibility of the formation of anion-radical oxygen participating in the formation of silver oxides. 

The results clearly indicate that in all considered clusters with stoichiometry of AgO the oxygen centers are radical-like. The spins on oxygen are ferromagnetically ordered for small clusters, while for the larger clusters antiferromagnetic state is likely to be more favourable. 

The spin density and charge on radical-like oxygen centers in the (AgO)$_\mathrm{n}$ particles is larger than that in Ag$_2$O cluster. 

The anion-radical oxygen is suggested to have the same nature with atomic electrophilic oxygen species which govern silver-catalized ethylene epoxidation and the oxygen form in the oxide AgO.\\

\textbf{Acknowledgments}\\

This work was supported by the Siberian Supercomputer Center SB RAS intergration grant no. 130 and the Russian Foundation for Basic Research (project no. 10-03-00441). One of us (E.V.T.) would like to thank Dr. S.Ph. Ruzankin and Dr. V.V. Kaichev for helpful discussions and International Charitable Zamaraev Foundation for partial financial support.

\bibliographystyle{model1a-num-names}
\bibliography{AgO}

\end{document}